\documentclass{article}

\usepackage{fontspec}
\usepackage{arxiv}
\renewcommand{\shorttitle}{Bridging Psychometric and Content Development Practices with AI}
\usepackage{hyperref}
\usepackage{url}
\usepackage{booktabs}
\usepackage{amsfonts}
\usepackage{nicefrac}
\usepackage{microtype}
\usepackage{graphicx}
\usepackage{natbib}
\usepackage{doi}

\title{Bridging Psychometric and Content Development Practices with AI: A Community-Based Workflow for Augmenting Hawaiian Language Assessments}

\author{
  Pōhai Kūkea-Shultz \\
  Kaiapuni Assessment of Educational Outcomes (KĀʻEO) Program\\
  University of Hawaiʻi at Mānoa College of Education\\
  \AND
  Frank Brockmann \\
  Center Point Digital Solutions\\
  (contracted to Research Corporation of University of Hawaiʻi)\\
  \texttt{frank@centerpointcorp.com} \\
}

\date{December 14, 2025}

\begin{document}

\maketitle

\begin{abstract}
This paper presents the design and evaluation of a community-based artificial intelligence (AI) workflow developed for the Kaiapuni Assessment of Educational Outcomes (KĀʻEO) program, the only native language assessment used for federal accountability in the United States. The project explored whether document-grounded language models could ethically and effectively augment human analysis of item performance while preserving the cultural and linguistic integrity of the Hawaiian language.
\medskip

Operating under the KĀʻEO AI Policy Framework, the workflow used NotebookLM for cross-document synthesis of psychometric data and Claude 3.5 Sonnet for developer-facing interpretation, with human oversight at every stage. Fifty-eight flagged items across Hawaiian Language Arts, Mathematics, and Science were reviewed during Round 2 of the AI Lab, producing six interpretive briefs that identified systemic design issues such as linguistic ambiguity, Depth-of-Knowledge (DOK) misalignment, and structural overload.
\medskip

The findings demonstrate that AI can serve as an ethically bounded amplifier of human expertise, accelerating analysis while simultaneously prioritizing fairness, human expertise, and cultural authority. This work offers a replicable model for responsible AI integration in Indigenous-language educational measurement.
\end{abstract}


\section{Introduction}

The Kaiapuni Assessment of Educational Outcomes (KĀʻEO) program is the only large-scale, Hawaiian-medium assessment system in the world and the only native language assessment used for federal accountability in the United States. It serves students in Hawaiian language immersion schools who are taught entirely through Hawaiian, and it operates within a community that views language as both a cultural right and a sacred trust \citep{KūkeaShultz2021, KūkeaShultz2023, Englert2025, KūkeaShultz2025}. Because all test content must remain in Hawaiian, every stage of item development from writing to psychometric validation occurs within a protected linguistic environment. This commitment to linguistic sovereignty, while essential, also creates a structural challenge: most psychometric analysts are not fluent in Hawaiian, and requiring content developers to translate their work into English solely to enable psychometric review places an unnecessary burden on a small team already balancing technical, linguistic, and cultural demands. As a result, systematic analysis of item quality has historically been limited, depending heavily on bilingual experts who could bridge both domains under tight resource constraints.

To address this persistent gap, the KĀʻEO project team established an AI Lab to investigate whether carefully governed artificial-intelligence tools could ethically and fairly augment the process of item analysis and interpretation. The emergence of large language models (LLMs) in educational contexts has generated considerable interest in automated item generation and analysis \citep{Hao2024, Wang2024}, with recent research demonstrating that LLMs can generate psychometrically acceptable assessment items when properly validated \citep{Gotz2023, Lee2024}. However, the critical question for Indigenous language contexts remains whether such tools can function within frameworks that prioritize community sovereignty and cultural integrity \citep{Lewis2020indigenous, Carroll2020}.

The goal was not to automate human judgment, but to build a workflow that could bridge the technical and linguistic divides separating psychometric data from content development practice. Specifically, the team sought to determine whether document-grounded large-language-model systems could:

\begin{enumerate}
    \item Interpret psychometric indicators (difficulty, discrimination, differential item functioning) in relation to item content;
    \item Generate interpretable, developer-friendly explanations of why items did or did not function as intended; and
    \item Do so within a closed, culturally respectful environment that protects Hawaiian language authority and community ownership of assessment data.
\end{enumerate}

This paper presents the results of that inquiry. It documents the design and evaluation of a community-based, AI-augmented workflow for Hawaiian-medium assessment development that bridges linguistic and psychometric expertise across language boundaries. Section 2 describes the ethical and governance framework that guided the work. Section 3 outlines the methodological design of the AI Lab. Section 4 demonstrates the workflow in action through detailed case examples, followed by a discussion of implications for culturally governed measurement systems (Section 5) and concluding observations on the broader significance of AI-supported collaboration in Indigenous-language contexts (Section 6).

\section{Ethical and Governance Framework}

The foundation of the AI Lab rested on the \textbf{KĀʻEO AI Principles and Policy Framework (v4.0)}, which defines how artificial intelligence may be used in service of the KĀʻEO program. The policy articulates seven non-negotiable values: stewardship of student data, protection of linguistic integrity, equitable access, appropriate integration with pedagogy, transparent communication about AI's function, attention to environmental impact, and the primacy of human expertise over technological efficiency. These principles formed the operational boundaries for every phase of the project, ensuring that the introduction of automation would enhance, rather than diminish, community authority over the assessment process.

This framework aligns with emerging Indigenous data sovereignty principles, particularly the CARE Principles for Indigenous Data Governance \citep{Carroll2020}, which emphasize Collective Benefit, Authority to Control, Responsibility, and Ethics in data management. The Indigenous Protocol and Artificial Intelligence Position Paper \citep{Lewis2020indigenous} further establishes that Indigenous communities must maintain decision-making authority over AI systems that engage with their cultural knowledge and languages.

All project activities occurred within a \textbf{closed analytic environment.} The NotebookLM workspace operated under a Business account with ephemeral data storage; session histories were automatically deleted upon closure, and no information was transmitted to external model-training pipelines. Item data were encrypted and accessible only to authorized analysts working under the KĀʻEO project team. This structure upheld the program's data sovereignty mandate, which specifies that Hawaiian assessment data remain under community control and never used for commercial or external research purposes.

Equally central was the \textbf{human-in-the-loop requirement,} although the AI Lab reframed this principle: humans are not merely checkpoints within an automated process; they are the loop. The policy prohibits unreviewed AI outputs, mandating that all analytic results be verified by qualified educators and psychometric experts before any recommendation is accepted. Recent research on human-AI collaboration emphasizes that optimal outcomes in high-stakes domains require maintaining meaningful human control rather than full automation \citep{Natarajan2024, Hao2024}. In practice, this meant that the AI provided a structured hypothesis about item performance, which was then interrogated by human reviewers who brought cultural, linguistic, contextual, and technical expertise to the interpretation. The process explicitly recognized that AI systems, while capable of pattern recognition, lack cultural and contextual understanding, exhibit sycophantic tendencies, and may produce spurious inferences when faced with ambiguity.

The framework also required that each AI initiative include a formal project plan approved by the KĀʻEO team, outlining purpose, benefits, potential weaknesses, and sociocultural impacts before implementation. High-stakes applications, such as those affecting operational test items, were limited to research phases until verified by external review. Periodic audits and written documentation of each decision maintained transparency and accountability.

Finally, the policy acknowledged the environmental footprint of large-scale computation, requiring that the educational value of AI use justify its energy cost. By treating ethical oversight and sustainability as design parameters rather than afterthoughts, the framework ensured that technological innovation remained subordinate to cultural responsibility. This governance structure made it possible for the AI Lab to explore new analytic methods while honoring the principle that technology must serve—and never supplant—the goals of the KĀʻEO program. It established the moral architecture within which the subsequent methodological work could proceed.

In addition to this framework, fairness in the application of AI tools was considered. According to the 2014 Standards, fairness is defined as ``responsiveness to individual characteristics and testing contexts so that test scores will yield valid interpretations for intended uses'' and also recognizes that ``fairness is a fundamental validity issue and requires attention throughout all stages of test development and use'' \citep[p.~49--50]{AERA2014}. This framework enabled the project to prioritize fairness, not just as it relates to the unique characteristics of Hawaiian language learners, but also to the context of their learning and subsequent assessment of the learning. The framework also guided decisions about where AI could appropriately augment the test development process without compromising the program's high bar for technical quality and fairness.

\section{Methodology and Design of the AI Lab}

The AI Lab was organized as a design-based research (DBR) project, emphasizing iterative refinement over fixed experimentation. Its central hypothesis was that a document-grounded, large language model-based workflow could expand key functions of human psychometric review: pattern detection, cross-item comparison, and interpretive synthesis. Such a workflow could serve as an analytic bridge between psychometric analysis and content development expertise without compromising linguistic sovereignty or data security.

The use of document-grounded generation, also known as Retrieval-Augmented Generation (RAG), addresses a critical challenge in applying LLMs to specialized domains: ensuring that generated content remains factually grounded in authoritative sources rather than hallucinating plausible-sounding but incorrect information \citep{Lewis2020rag, Gao2024}. The workflow maintained traceability by restricting the model's reasoning to explicitly ingested KĀʻEO documents: psychometric reports, item development guidelines, flagged and exemplar items, style guides, specifications, and technical standards such as the AERA/APA/NCME's \textit{Standards for Educational and Psychological Testing}. Synthesizing this range of sources manually is impractical during routine item development; the workflow automated the aggregation while preserving human interpretive authority, and the use of document-grounded generation helped prevent the introduction of external or fabricated claims.

A critical design principle was respecting the time and expertise of Hawaiian-language item writers. These content developers are experts in language and pedagogy, not psychometrics. Reviewing extensive statistical reports and synthesizing patterns across dozens of flagged items is not the best use of their time. The AI workflow was designed to handle that initial aggregation, identify potential issues related to item development best practices, and perform pattern recognition across the flagged set. This approach honored the cultural and linguistic authority of item writers by presenting them with interpretable, actionable summaries for their review rather than raw data, allowing them to focus their time on what only they can do: making culturally informed decisions about language and content.

\subsection{Iterative Development}

The Lab evolved through two cycles:

\begin{itemize}
    \item \textbf{Round 1 (Pilot, August and September 2025):} A small-scale feasibility test that established the ethical review process, prompt templates, and reporting formats. Outputs from this round became the ``Lab Docs,'' the standard operating instructions for all later analyses.
    \item \textbf{Round 2 (Full Implementation, October 2025):} Expanded the analysis to three content areas: Hawaiian Language Arts (HLA), Mathematics, and Science. This round introduced full psychometric datasets and the complete suite of reference materials described in the AI Lab Project Instructions. Round 2 ultimately generated the six developer briefs that form the evidentiary base of this paper.
\end{itemize}

\subsection{Data Corpus}

Across content areas, the Lab examined \textbf{58 flagged assessment items:} 12 in HLA, 27 in Mathematics, and 19 in Science. Each record contained item metadata (Student Learning Objective [SLO], Depth of Knowledge [DOK], item type), psychometric indicators (difficulty p-value, point-biserial correlation, fit statistics, differential-item-functioning [DIF] flags), and qualitative review notes from prior development cycles. For every flagged item, 3–5 high-performing exemplars served as comparators, enabling cross-item benchmarking of linguistic clarity, cognitive load, and construct alignment.

\subsection{Analytic Focus}

The analytic questions guiding Round 2 were pragmatic:

\begin{enumerate}
    \item Can a grounded-AI system reliably trace the technical causes of poor item performance?
    \item Can it generate interpretable guidance for item developers?
    \item Can it do so without leaking or mistranslating culturally sensitive content?
\end{enumerate}

The objective was methodological demonstration rather than inferential generalization. Success was measured by completeness, accuracy, and human acceptance of each analytic report.

\subsection{Human Roles and Review Structure}

The KĀʻEO lead psychometrician provided the foundational psychometric analysis: flagged items, statistical indicators, and technical documentation that served as input to the AI Lab process. Two Lab operators conducted prompt submissions, document ingestion, and environment configuration. Two reviewers then validated AI outputs: the lead Lab operator verified that AI-generated interpretations accurately reflected the underlying psychometric data, while a member of the KĀʻEO team assessed cultural and linguistic fidelity. The KĀʻEO team, who possessed necessary cultural and linguistic knowledge, served as the final gate; no AI-generated guidance reached content developers without their approval.

\subsection{Toolchain Overview}

The Lab employed a two-model configuration:

\begin{itemize}
    \item \textbf{NotebookLM (Google, Business Workspace):} Provided document-grounded reasoning within an ephemeral, closed environment and Zero Data Retention (ZDR) prompting. Approximately 50–70 source documents per content area were ingested, including technical manuals, style guides, and exemplar items.
    \item \textbf{Claude 3.5 Sonnet (Anthropic):} Used in structured-output rendering mode (artifact generation) to translate technical syntheses into developer-readable interpretive briefs. All processing occurred in sessions with Anthropic's standard 30-day retention policy with opt-out from contributing to training data; to comply with data-sovereignty rules, Claude had no direct access to secure test materials and was specifically tasked with ensuring the NotebookLM outputs were compliant with the guidance in the Guidelines for Creating Developer-Focused Interpretive Briefs v 4.0 instructions.
\end{itemize}

\subsection{Evaluation and Refinement}

Outputs were evaluated against three criteria: (1) \emph{evidence discipline}: every analytic claim had to cite an explicit source document; (2) \emph{linguistic integrity}: Hawaiian text could be paraphrased but never altered or publicly translated; and (3) \emph{cultural fidelity}: the tone and framing had to align with Hawaiian educational values. Feedback from reviewers in Round 2 led to adjustments in prompt structure and to the integration of explicit ``deference notes'' accompanying all AI-generated Hawaiian drafts, signaling that such text served only as provisional models for expert refinement.

\subsection{Transition}

By the conclusion of Round 2, the Lab had produced a complete, reproducible workflow linking psychometric data, linguistic analysis, and culturally governed human review. The next section documents this workflow in action and evaluates its performance through representative case studies drawn from the Hawaiian Language Arts and Mathematics datasets.

\section{Results: Demonstrating the AI-Augmented Workflow}

\subsection{Workflow Overview}

\begin{figure}[ht]
    \centering
    \includegraphics[width=0.9\textwidth]{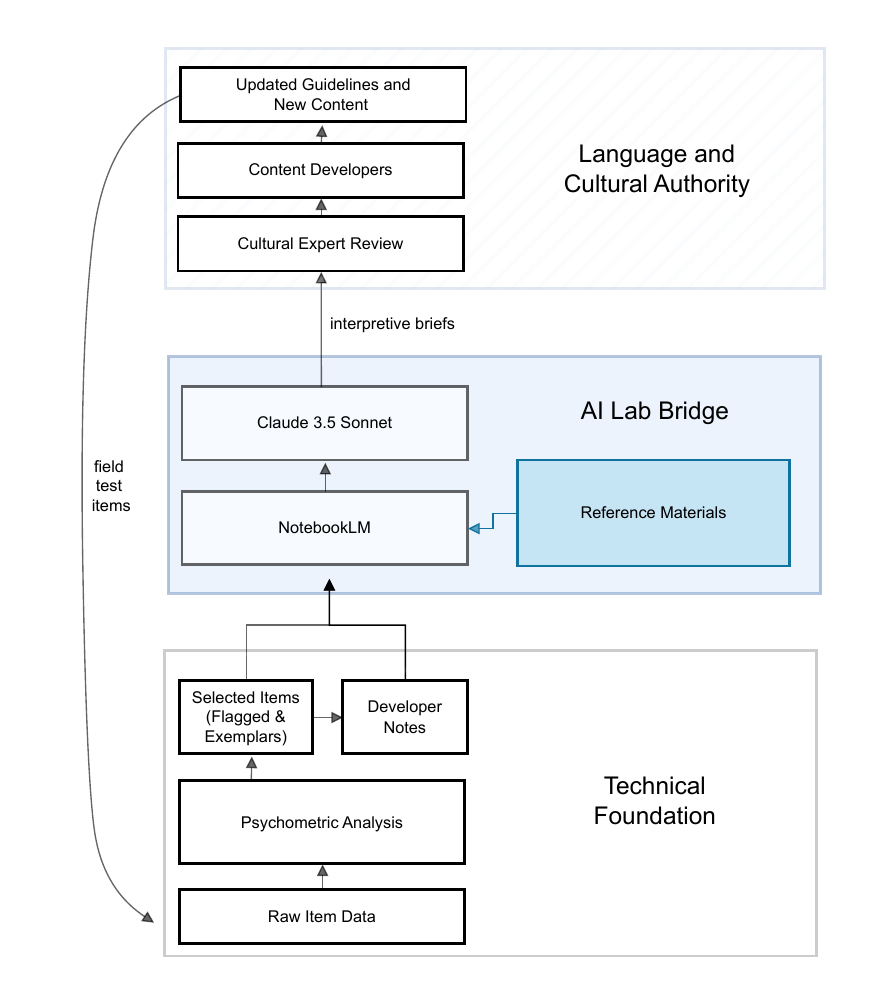}
    \caption{\textit{AI-augmented workflow for KĀʻEO item analysis, illustrating that humans \emph{\textbf{are}} the loop rather than merely \emph{\textbf{in}} it. Psychometric data, exemplar items, and developer notes feed into a document-grounded AI synthesis process (NotebookLM, Claude 3.5 Sonnet), with cultural expert review serving as a gate before outputs reach content developers. The feedback loop shows how revised guidelines inform future item development.}}
    \label{fig:workflow}
\end{figure}

The Round 2 AI Lab review was designed to test whether a document-grounded language-model workflow could perform structured, replicable analyses of Hawaiian-medium assessment items within the parameters of the KĀʻEO AI Policy Framework. The system was configured as a three-tier process: (1) AI-assisted item diagnostics in NotebookLM, (2) cross-item synthesis and pattern recognition, and (3) developer-facing narrative translation through Claude 3.5 Sonnet. Figure~\ref{fig:workflow} illustrates this workflow structure.

All work occurred between 10–17 October 2025 in a Google NotebookLM Business Workspace operating in ephemeral mode, which by default deletes all conversational history after each session. The environment was closed to the public web, and every notebook instance was linked to encrypted internal storage consistent with the KĀʻEO data-sovereignty protocols. Approximately 50–70 source documents were ingested across content areas (Hawaiian Language Arts, Mathematics, and Science), including item-level psychometric summaries, the KĀʻEO Item Development Manual, technical reports, and bias-sensitivity rubrics.

The analytic design followed the \textbf{AI Lab Frameworks for Reviewing KĀʻEO Items (v1.7)}, which requires strict data attribution and prohibits fabricated or unsupported inferences. Each problematic item entered the system with its complete metadata (SLO code, DOK assignment, difficulty, item type, psychometric flags). NotebookLM performed the initial structured analysis, generating an evidence-linked Synthesis Report for each item. Human reviewers validated every flagged issue before authorizing cross-item aggregation.

\subsection{The Interpretability Challenge}

Historically, psychometric review of KĀʻEO items has been constrained by a structural gap: the assessment is written entirely in Hawaiian by fluent content developers, but the lead psychometrician and technical analysts are not Hawaiian speakers. Conversely, content developers are experts in language and pedagogy, not psychometrics.

This division of expertise has meant that item-level statistical flags (low discrimination, unexpected difficulty, DIF) could be identified, but interpreting why an item performed poorly required either translation of secure content (violating confidentiality protocols) or reliance on a very small number of bilingual intermediaries who could bridge both domains. The result was a critical tension: psychometric patterns were visible, but their causes in item structure, wording, or cultural framing remained opaque to those best positioned to address them. This challenge is not unique to Hawaiian-medium assessment; understanding why statistically flagged items underperform is a persistent challenge in educational measurement \citep{Angoff1993}, particularly in pinpointing structural issues or patterns that might be present across multiple items.

The AI Lab workflow was designed to solve this interpretability problem by creating a secure analytical space where content could be examined across languages without being publicly translated, and where technical and linguistic reasoning could be synthesized into actionable developmental guidance.

Because all KĀʻEO assessment items are written exclusively in Hawaiian, traditional psychometric review processes have historically excluded non-fluent analysts. The AI Lab workflow provided, for the first time, a secure and ethical mechanism for interpretable cross-language analysis. Within the closed NotebookLM environment, model-generated English paraphrases were treated solely as temporary analytic artifacts, allowing psychometric reasoning about each item's structure and technical characteristics while keeping Hawaiian source text authoritative and unaltered.

\subsection{Tool Configuration and Rationale}

Two complementary models were used to balance technical rigor and human readability.

NotebookLM was selected for its ability to conduct document-grounded reasoning—that is, to restrict the model's claims to information explicitly present in the ingested corpus. This feature allowed the Lab to trace each analytic statement back to a verifiable source document, fulfilling the ``strict evidence discipline'' clause of the governing framework. Its structured-prompt environment also supported multi-document reasoning that approximates a traditional psychometric audit but at a scale no individual analyst could complete manually.

Claude 3.5 Sonnet (Anthropic, structured-output rendering mode) served as the final language-refinement engine. Using the technical outputs from NotebookLM, Claude produced developer-friendly interpretive briefs that aligned with the \textbf{Guidelines for Creating Developer-Focused Interpretive Briefs (v4.0)}. This ``editorial rinse'' step transformed dense analytic text into accessible narratives without introducing new technical claims.

Human oversight followed the structure described in Section 3.4: Lab operators handled AI configurations, while reviewers verified outputs against psychometric accuracy and cultural integrity criteria, with KĀʻEO staff exercising final authority. This layered configuration maintained human authority over interpretation while demonstrating how an ethically governed AI workflow can augment the analytical capacity of a small, specialized Hawaiian-medium assessment team.

This arrangement effectively bridged the long-standing linguistic divide between psychometric analysis and content expertise. It enabled collaboration between non-Hawaiian-speaking technical analysts and fluent Hawaiian educators without requiring public translation or data export, demonstrating that responsible AI use can expand—not replace—human expertise in linguistically sovereign assessment programs.

\subsection{Process Demonstration: Cross-Item Analysis in Hawaiian Language Arts}

To illustrate how the AI-augmented workflow functioned in practice, the team focused on two Grade 3 Hawaiian Language Arts items that exhibited persistently anomalous performance. Both had extremely low discrimination indices (0.122 and 0.03, respectively) despite appearing structurally sound. The AI's initial analysis indicated that each item required students to distinguish between near-synonymous Hawaiian expressions differentiated only by subtle contextual conventions or grammatical rules such as verb transitivity. These fine-grained lexical distinctions appeared to have repeatedly challenged even high-performing students, producing inverse discrimination patterns.

\subsection{AI-Assisted Synthesis and Pattern Recognition}

Using the NotebookLM environment, the operators ingested the relevant reference materials (psychometric summaries, the KĀʻEO Item Development Manual, and linguistic guidance documents) alongside the item metadata. NotebookLM's cross-item reasoning was tasked with identifying whether the anomalies shared a structural cause. Within minutes, the system produced a synthesis noting that both items depended on ``fine-grained lexical distinctions not reliably taught across cohorts.'' This observation would have required manually reviewing dozens of spreadsheets and item templates; the AI's ability to aggregate this evidence rapidly demonstrated its analytic efficiency.

\subsection{Translation and Human Validation}

The AI's draft analysis was exported locally and then uploaded to Claude 3.5 Sonnet, which generated a developer-friendly interpretive brief summarizing the shared pattern as \emph{Linguistic Ambiguity and Fine Grammatical Traps}. Claude reframed psychometric jargon (e.g., negative item-rest correlation) into accessible language describing ``unexpected patterns where higher-performing students selected incorrect answers.'' Human reviewers then assessed these interpretations: the lead Lab operator verified that the statistical claims were consistent with the lead psychometrician's original technical-report data, while KĀʻEO staff examined the Hawaiian exemplars for cultural and linguistic appropriateness and alignment. No brief advanced to content developers until KĀʻEO staff confirmed its cultural and practical appropriateness.

\subsection{Limits of AI Interpretation}

KĀʻEO staff feedback underscored a critical boundary of the approach: they noted that NotebookLM's interpretation of certain grammatical nuances lacked the figurative depth and layered meaning characteristic of Hawaiian, remarking that ``lack of Hawaiian language proficiency or the ability to understand the nuance and highly figurative nature of Hawaiian may be impacting the analysis.'' This insight confirmed that while the AI could map where items failed psychometrically, it could not necessarily explain why those failures occurred within the cultural logic of the language.

The items discussed in Section 4.4 illustrate a deeper issue: large language models built primarily on English-language training data cannot fully comprehend the cultural and worldview differences embedded in other languages. In this case, the AI made assumptions about grammatical constructions that might be considered errors in English, but are highly valued language skills in Hawaiian. KĀʻEO staff identified this misinterpretation during review; while documented in the briefs, it was excluded from the training sessions delivered to content developers. Left unchecked, such misinterpretations could introduce fairness concerns by failing to account for the values and worldview of Hawaiian language speakers and learners.

This limitation reinforces why humans as the loop (rather than merely in the loop) remained essential. Without cultural expertise serving as a gate, LLM training data shortcomings risk inaccurately shaping interpretations, item development guidance, and ultimately introduce fairness concerns into the use and interpretation of the test itself. The AI's pattern recognition, however efficient, still required human judgment to become defensible and actionable guidance for item development.

\subsection{Human Review and Validation}

All AI-generated interpretive briefs were evaluated against three criteria: (1) psychometric accuracy, (2) cultural and linguistic fidelity, and (3) actionable clarity for content developers. During Round 2, six briefs were produced across three content areas; all six passed review without requiring fundamental revision. In at least one case, AI-generated guidance that mischaracterized Hawaiian grammatical features was excluded from the training materials (see Section 4.7).

Formal validation metrics (acceptance rates, inter-rater agreement, revision frequency) remain a target for future research as the workflow scales. The present study establishes the feasibility and procedural integrity of the approach; systematic outcome evaluation will require larger item samples and structured feedback protocols across multiple development cycles.

No AI-generated guidance was released to developers without approval from both reviewers, with KĀʻEO staff holding final authority on cultural and practical appropriateness. This validation structure ensured that the workflow's efficiency gains did not compromise the quality or cultural and practical appropriateness of the final analytic product.

\subsection{Outcome and Implications}

The combined review yielded some actionable models for future item writing. For example, a key finding across Mathematics items was an effect the briefs termed ``structural overload'': items assigned DOK 1 or 2 that actually required students to perform multiple sequential calculations before selecting a single answer, effectively ``stacking'' several simpler tasks into one. The AI-generated briefs recommended redesigning such items using a Multi-Part (MP) format, which scaffolds complexity across linked steps and ensures that performance issues in later parts reflect strategic reasoning challenges rather than earlier errors in computation.

These recommendations were integrated into the Round 2 developer guidelines and later endorsed for use in follow-up training sessions. More importantly, the process demonstrated that AI, used under human cultural oversight, can extend analytical reach without supplanting expertise. The workflow effectively bridged the structural and linguistic divide between psychometric evaluation and content development to offer a scalable, ethically governed model for Hawaiian-medium assessment innovation.

\section{Discussion: Implications for AI-Augmented Measurement in Cultural Contexts}

The Round 2 study demonstrated that an AI-assisted, document-grounded workflow can operate within the constraints of a culturally governed assessment system without eroding human authority. What the process revealed most clearly is that speed and scale alone do not constitute insight; sound, defensible interpretation still requires disciplined human judgment. The KĀʻEO workflow succeeded not because the models were autonomous, but because they were carefully bounded: ephemeral, evidence-linked, and embedded in an existing professional culture of review.

\subsection{Methodological Innovation and Human Control}

In traditional psychometric review, the technical and linguistic dimensions of item quality are handled by different experts, often separated by both geography and language. By introducing NotebookLM and Claude into this structure, we inverted the conventional framing: rather than placing humans in an AI-driven loop, we embedded AI tools within a human-governed workflow. With their cultural, linguistic, and psychometric expertise, the human constituents of the AI Lab remained not merely part of the workflow, but the loop itself. The AI Lab's design thus re-defined ``human-in-the-loop'' as multi-expert collaboration through AI mediation.

This approach expands analytical capacity for Indigenous-language programs, language revitalization initiatives, and other multilingual educational contexts that cannot afford large psychometric or linguistic teams. Research on culturally responsive assessment for Indigenous students emphasizes the importance of community involvement and Indigenous epistemologies \citep{Preston2021, IES2025, KūkeaShultz2023}, principles that guided the workflow design. It also reframes AI not as a scorer or author, but as a workflow component acting as a specialized assistant embedded in a chain of accountable humans. The goal is to extend the reach of both cultural-linguistic and psychometric expertise, not to replace either.

Notably, the Round 2 results clarified where that extension was most effective. AI excelled at identifying technical and structural patterns: DOK misalignment, cognitive overload, format constraints that introduced construct-irrelevant variance. These are domains where pattern recognition across large document sets provides clear efficiency gains. By contrast, AI output was less useful in areas where cultural and linguistic judgment was required; it misinterpreted some Hawaiian grammatical features as errors, missed figurative depth, and produced interpretations that lacked the relational grounding essential to Hawaiian understanding and ways of knowing. This outcome suggests a practical division: use AI for structural and technical synthesis, but keep cultural and linguistic interpretation in human hands. Whether future LLMs might perform better in such contexts remains an open question; deeper representation of Indigenous languages in training data might improve surface-level comprehension, but the ability of algorithmic systems to grasp the relational, figurative, contextual and worldview nuances embedded in languages like Hawaiian is far from certain.

This also raises a broader point about fairness: the question is not only whether AI outputs are fair, but whether AI tools can operate fairly given their foundational limitations. Recognizing these constraints early does not preclude using AI; rather, it enables test developers to draw appropriate boundaries around AI contributions while honoring human expertise and upholding fairness.

\subsection{Ethical and Cultural Safeguards}

Operating under the KĀʻEO AI Policy Framework v 4.0, every step maintained the principles of data sovereignty, human oversight, and cultural deference. The workflow proved that closed-system analysis, combined with ephemeral data handling, could preserve confidentiality for live assessment content while still enabling technical research. More importantly, the inclusion of cultural review (embodied by the KĀʻEO staff's critique of algorithmic shallowness) ensured that the linguistic integrity of Hawaiian language remained the final standard. The AI's inability to interpret figurative depth was not a flaw to fix but a boundary to respect, reminding us that sound measurement practices in Indigenous contexts depend on relational meaning and cultural and community aspirations, not just statistical correlation \citep{KūkeaShultz2021, KūkeaShultz2023}.

\subsection{Strengths, Risks, and Transferability}

The principal strength of this workflow lies in its replicability across linguistically sovereign programs and under-resourced multilingual assessment contexts. Its modular design (secure data ingestion, AI synthesis, human verification, and interpretive translation) could be adapted for other languages or disciplines. However, several risks must be managed:

\begin{enumerate}
    \item \textbf{Over-trust in AI objectivity.} Without explicit evidence tracing, model output can appear authoritative when it is merely plausible.
    \item \textbf{Cultural drift.} Each iteration must be reviewed by fluent speakers and cultural experts to prevent subtle erosion of linguistic authenticity.
    \item \textbf{Environmental cost.} Large-model inference consumes energy; usage must remain proportional to educational benefit, consistent with Policy Clause 6 on environmental stewardship.
\end{enumerate}

These risks underscore that governance, not technology, determines whether AI strengthens or weakens educational systems.

\subsection{Reframing Psychometric Practice}

The experiment also invites a broader re-definition of what counts as psychometric inquiry. Historically, ``analysis'' has meant quantitative modeling of item responses. Here, it also meant qualitative reasoning about cultural fit, accessibility, and construct clarity—all mediated through AI. By formalizing this mixed-methods workflow, the KĀʻEO program contributes to a growing field of \textbf{computational measurement design} \citep{Hao2024}, where language models serve as cognitive partners that map complex qualitative domains into interpretable analytic form. For programs operating within multilingual technical teams similar to this Hawaiian-medium assessment program, the model represents a fundamental shift: measurement quality and fairness now depend as much on ethical infrastructure as on statistical evidence, with an operational workflow designed to both serve and amplify human psychometric judgment rather than replacing it. In the same way that KĀʻEO has pushed the field of measurement to consider expanding its view of validity to include concepts of cultural and community validity \citep{KūkeaShultz2021, KūkeaShultz2023, KūkeaShultz2025, Englert2025}, utilizing AI tools while also adhering to the cultural and community validity ideals that support fairness is a logical next step. As the measurement community grapples with the use of AI in test development, interpretation and use and how fairness is impacted (or ignored) by these tools, this AI lab (and its closely-held policy framework) can serve as a roadmap for navigating the rapid adoption of AI tools in testing.

\subsection{Future Directions}

The next logical step is systematic evaluation: examining both psychometric outcomes (whether revised items exhibit improved discrimination and reduced DIF) and process outcomes (how the workflow affects content developer understanding, efficiency, and confidence in item construction). Comparing developer acceptance of AI-generated briefs against traditional psychometric feedback would also illuminate whether the interpretive bridge functions as intended. Parallel work could explore integration with translation-aware LLMs fine-tuned on Hawaiian corpora under local stewardship. However, any expansion must remain anchored to the original KĀʻEO principles: human oversight, linguistic integrity, and cultural accountability.

\section{Conclusion}

The AI Lab initiative demonstrated that technology, when governed by community ethics and human expertise, could responsibly extend the reach of psychometric analysis into domains once considered inaccessible. By bridging psychometric and content-development practices through an AI-assisted workflow, the KĀʻEO program transformed the analytic process from a siloed, language-bounded activity into a collaborative cycle linking data, design, and cultural interpretation.

The system functioned not through automation but through the structure that surrounded it: secure environments, document-grounded reasoning, and cultural oversight that preserved the integrity of Hawaiian language while enabling transparent technical review. Throughout, humans remained the loop itself, not merely participants within an automated process. What emerged is a replicable model of \textbf{augmentative intelligence} \citep{Hao2024, Natarajan2024}, one that honors human judgment while amplifying analytical capacity.

In sum, the AI Lab project demonstrated that ethically bounded language models can function as powerful amplifiers of human expertise within Indigenous and Hawaiian-medium educational measurement, while leveraging policy guidelines that ensure technical quality throughout the process. They do not replace the cultural or psychometric wisdom of experts; they extend its reach, creating the possibility of sustainable, community-governed innovation in how we understand assessment quality, improve practice, and prioritize fairness.

\section*{Acknowledgments}

The authors gratefully acknowledge the Kaiapuni Assessment of Educational Outcomes (KĀʻEO) program and the Hawaiʻi State Department of Education for their support of this research. We extend particular thanks to Christopher Estep for his contributions to the AI Lab operations and technical configuration; and Kerry Englert, whose psychometric analysis provided the empirical foundation upon which the AI Lab process was built.

Frank Brockmann's work on this study was performed under contract to the Hawaiʻi State Department of Education's KĀʻEO Program via the Research Corporation of University of Hawaiʻi.

This work was conducted in accordance with the KĀʻEO AI Principles and Policy Framework (v4.0), which prioritizes cultural sovereignty, data stewardship, and the primacy of human expertise in all AI-assisted educational measurement activities.

\bibliographystyle{apalike}
\bibliography{references}

\end{document}